\documentclass[twocolumn]{aastex631} 
\usepackage{gensymb}
\usepackage{graphicx}
\usepackage{amssymb}

\newcommand{\hi}{\ion{H}{1}}    \newcommand{\oi}{\ion{O}{1}}  
\newcommand{\few}{\ion{Fe}{2}}  \newcommand{\siw}{\ion{Si}{2}} 
\newcommand{\alw}{\ion{Al}{2}}  \newcommand{\sw}{\ion{S}{2}} 
\newcommand{\no}{\ion{N}{1}}    \newcommand{\fet}{\ion{Fe}{3}}
    \newcommand{\kms}{\,km\,s$^{-1}$}
 \newcommand{\tn}{\tablenotetext}

\received{February 22, 2023}
\revised{March 14, 2023}
\accepted{March 21, 2023}
\submitjournal{ApJL}
\shorttitle{Dust in Complex C}
\shortauthors{Fox, Cashman, Kriss et al.}

\begin{document}

\title{Detection of Dust in High-Velocity Cloud Complex C -- Enriched Gas Accreting onto the Milky Way
\footnote{Based on observations made with the NASA/ESA Hubble Space Telescope, obtained from the Data Archive at the Space Telescope Science Institute, which is operated by the Association of Universities for Research in Astronomy, Inc., under NASA contract NAS5-26555. These observations are associated with program 16196.}}
\correspondingauthor{Andrew Fox}
\email{afox@stsci.edu}

\author[0000-0003-0724-4115]{Andrew J. Fox}
\affil{AURA for ESA, Space Telescope Science Institute, 3700 San Martin Drive, Baltimore, MD 21218}
\affil{Department of Physics \& Astronomy, Johns Hopkins University, 3400 N. Charles St., Baltimore, MD 21218}
\author[0000-0003-4237-3553]{Frances H. Cashman}
\affiliation{Space Telescope Science Institute, 3700 San Martin Drive, Baltimore, MD 21218, USA}
\author[0000-0002-2180-8266]{Gerard A. Kriss}
\affiliation{Space Telescope Science Institute, 3700 San Martin Drive, Baltimore, MD 21218, USA}
\author[0000-0003-3242-7052]{Gisella de Rosa}
\affiliation{Space Telescope Science Institute, 3700 San Martin Drive, Baltimore, MD 21218, USA}
\author[0000-0002-2509-3878]{Rachel Plesha}
\affiliation{Space Telescope Science Institute, 3700 San Martin Drive, Baltimore, MD 21218, USA}
\author[0000-0002-0957-7151]{Yasaman Homayouni}
\affiliation{Space Telescope Science Institute, 3700 San Martin Drive, Baltimore, MD 21218, USA}
\affiliation{Department of Astronomy and Astrophysics, The Pennsylvania State University, 525 Davey Laboratory, University Park, PA 16802}
\affiliation{Institute for Gravitation and the Cosmos, The Pennsylvania State University, University Park, PA 16802}
\author[0000-0002-1188-1435]{Philipp Richter}
\affil{Institut f\"ur Physik und Astronomie, Universit\"at Potsdam, Haus 28, Karl-Liebknecht-Str. 24/25, D-14476, Potsdam, Germany}

\begin{abstract}
We present the detection of dust depletion in Complex C, a massive, infalling, low-metallicity 
high-velocity cloud in the northern Galactic hemisphere that traces the ongoing accretion of gas onto 
the Milky Way. We analyze a very high signal-to-noise HST/COS spectrum of AGN \object{Mrk\,817} formed by 
coadding 165 individual exposures taken under the AGN STORM 2 program, allowing us to determine 
dust-depletion patterns in Complex C at unprecedented precision.
By fitting Voigt components to 
the \oi, \sw, \no, \siw, \few, and \alw\ absorption and applying ionization corrections 
from customized \textsc{Cloudy} photoionization models, we find sub-solar elemental abundance ratios of 
[Fe/S]=--0.42$\pm$0.08, [Si/S]=--0.29$\pm$0.05, and [Al/S]=--0.53$\pm$0.08. 
These ratios indicate the depletion of Fe, Si, and Al into dust grains, since S is mostly undepleted. 
The detection of dust provides an important constraint on the origin of Complex C, as dust grains 
indicate the gas has been processed through galaxies, rather than being purely extragalactic.  
We also derive a low 
metallicity of Complex C of [S/H]=--0.51$\pm$0.16 ($\approx$31\% solar), 
confirming earlier results from this sightline.
We discuss origin models that could explain the presence of dust in Complex C, 
including Galactic fountain models,
tidal stripping from the Magellanic Clouds or other satellite galaxies, 
and precipitation of coronal gas onto dust-bearing ``seed" clouds.
\end{abstract}


\section{Introduction}\label{sec:intro}

Large-scale gas flows play essential roles in circulating mass, energy, momentum, metals, and dust grains 
between galaxies and their surroundings. The circulation of these basic physical quantities through galaxy 
halos drives and regulates galaxy evolution, with the rate of star formation ultimately controlled by the 
balance between the gas supply and the feedback that prevents it from collapsing into stars. Gas flows 
are therefore of paramount importance for galaxy evolution.

In the Milky Way (MW), we can dissect galaxy-scale gas flows in more detail than is possible in any other galaxy. 
Using spectroscopy, we can probe the physical, chemical, spatial, and kinematic properties of gas flows and 
how they distribute across the sky. Galactic inflows and outflows are typically traced via the high-velocity 
clouds (HVCs), clumps of multi-phase gas moving too fast for Galactic rotation, in practice meaning clouds 
with velocities in the Local Standard of Rest (LSR) reference frame of $|v_{\rm LSR}|>100$\kms\ 
\citep{wakker1997}. While HVCs were discovered and first characterized in neutral gas via \hi\ 21\,cm 
emission, most of their mass is in ionized gas, as revealed by UV absorption-line studies 
\citep[e.g.][]{lehner2012, richter2017, fox2019}.

\begin{figure*}[!ht]
\includegraphics[width=0.98\textwidth]{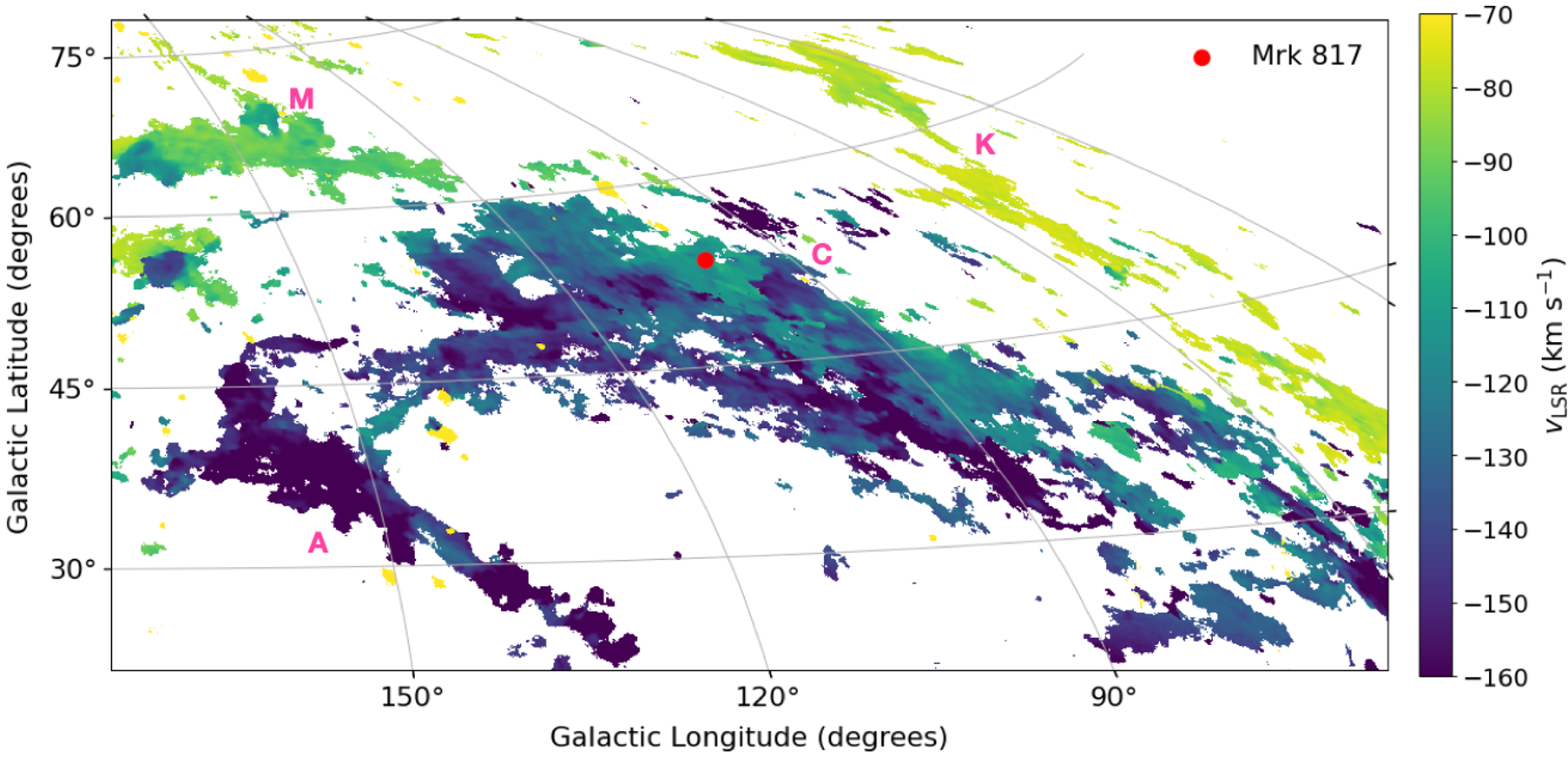}
\includegraphics[width=\textwidth]{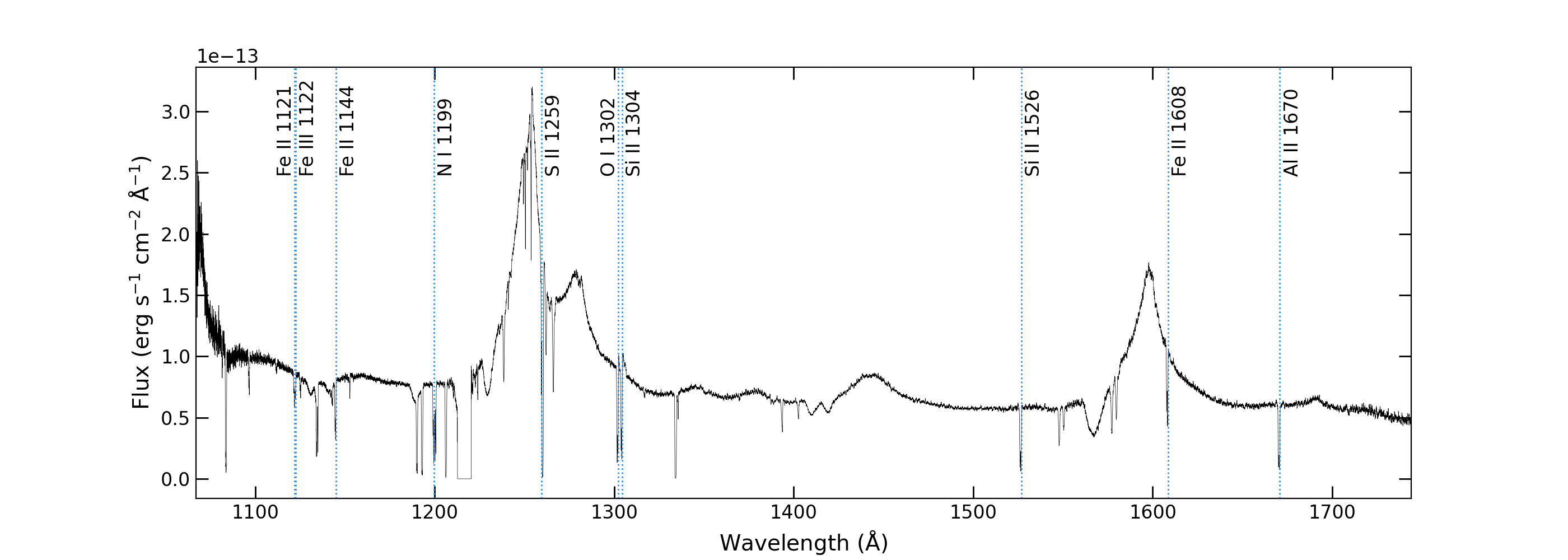}
\caption{{\bf Top:} \hi\ 21\,cm emission map of Complex C in Aitoff projection based on HI4PI data \citep{westmeier2018}. 
The map is integrated over the velocity range $-175<v_{\rm LSR}<-50$\kms. 
The location of \object{Mrk\,817} is marked with a red circle, and major HVC complexes are labeled in pink.
{\bf Bottom:} full coadded COS/FUV spectrum of \object{Mrk\,817} labeled with MW lines used in this analysis. 
The extremely high S/N of this spectrum enables high-precision measurements of Complex C absorption.
MW Lyman-$\alpha$ absorption was not observed because the G130M/1222 grating setting places it in the gap between COS detector segments A and B.} 
\label{fig:map}
\end{figure*}

Complex C is among the best studied of all HVCs. With an infalling velocity of $v_{\rm LSR}=-130$\kms\ and an 
angular size of 1\,700 square degrees in the northern Galactic hemisphere, Complex C is a poster child of 
accreting gas onto the Milky Way (see Figure~\ref{fig:map}). A series of UV absorption-line analyses have 
found a low metallicity in Complex C of $\approx$10--30\% solar 
\citep{wakker1999, murphy2000, richter2001, gibson2001, tripp2003, collins2003, collins2007, shull2011}, 
which point toward an extragalactic origin, though models where Complex C represents the return flow of 
a condensing Galactic fountain cannot be excluded \citep{fraternali2015}.
The cloud is at a distance of 10--12 kpc \citep{wakker2007, thom2008}, 
contains cold \hi\ filaments \citep{marchal2021}, 
and is surrounded by an envelope of highly ionized gas indicating an interaction with the hot 
Galactic halo \citep{fox2004}.

While these earlier analyses all concluded that Complex C has a low metallicity, they did not have the 
sensitivity to demonstrate that Complex C contains dust grains. 
Dust is important because it represents a vital clue to a cloud's origin, with Galactic clouds expected 
to contain dust and extragalactic clouds generally expected not to. 
In this Letter, we present an extremely high S/N HST/COS spectrum of \object{Mrk\,817}, an AGN lying behind 
Complex C, and use a UV depletion analysis to show that Complex C does contain dust, with 
$\approx$0.3--0.5\,dex of depletion seen in the refractory elements Si, Al, and Fe. 
We present our data reduction procedures in Section~\ref{sec:data}, analysis techniques in 
Section~\ref{sec:analysis}, and discussion and conclusions in Section~\ref{sec:discussion}. 

\section{Observations and Data Handling}\label{sec:data}

\object{Mrk\,817} is a Seyfert 1 galaxy at $z_{\rm em}$=0.031455 
that by good fortune lies directly behind HVC Complex C. This alignment allows a precision 
analysis of the cloud's chemical composition. 
\object{Mrk\,817} was observed with the Cosmic Origins Spectrograph \citep[COS;][]{green2012} onboard HST 
165 times in a 15-month period beginning 2020 November 24 under the reverberation mapping program 
AGN STORM 2 (HST Program ID 16196). Each visit consisted of four 60s exposures with 
G130M/1222 (one at each FP-POS position), exposures of 175s and 180s  with G160M/1533 (at FP-POS 1 and 2, respectively), 
and two 195s exposures with G160M/1577 (at FP-POS 3 and 4). All data were taken at COS FUV Lifetime Position 4. 
While the AGN STORM 2 program was designed to study the temporal variation in the continuum and line emission 
from the AGN itself \citep{kara2021}, the spectra can be co-added to form a single spectrum of exquisite sensitivity, 
ideal for studies of the absorption in the MW halo and the foreground intergalactic medium. 

The individual COS exposures were reduced using version 3.4.1 of the {\tt calcos} data reduction pipeline. We then created two co-added spectra for use in our analysis. 
The first is a coadd of all 165 exposures. Full details of the coaddition steps are described in \citet{homayouni2023}.
In brief, we used a Python implementation of the IRAF \texttt{splice} algorithm to resample the individual exposures onto a uniform wavelength grid before coaddition. Individual pixels were weighted by the exposure time in the \texttt{DQ\_WGT} array. 
Error bars for each binned pixel were recalculated based on the total counts in each bin, 
assuming Poisson statistics with the low-count correction of \citet{gehrels1986}.
The second coadd was formed from the 108 spectra taken in HST orbital night time, defined as those exposures with header keyword \texttt{SUN\_ALT}$<$20$^\circ$. This night-only spectrum minimizes geocoronal emission (airglow) from the sunlit side of the Earth's atmosphere, which contaminates the \oi\ $\lambda$1302 and \siw\ $\lambda$1304 lines, both key lines for our metallicity analysis. 

The full coadded spectrum has an extremely high signal-to-noise ratio per pixel 
of 66 at 1300\AA\ and 94 at 1528\AA, 
corresponding to 161 and 230 per resolution element, respectively (assuming 6 pixels per resolution element).
These exquisite S/N levels are possible because of the sheer number of individual exposures, 
the use of 4 FP-POS positions per grating, and the non-repeatability in the COS mode-select mechanism, 
which means identical exposures fall at slightly different $x$-positions on the FUV detector (by up to 10 pixels).
With 165 visits and 4 FP-POS per grating, each wavelength samples about 660 different detector 
locations, which greatly ameliorates pixel-level flat-field structure. 
The night-only spectrum has a S/N per pixel of 50 at 1300\AA\ (123 per resolution element). 

We applied a heliocentric-to-LSR velocity correction of 11.7 \kms, 
appropriate for the Galactic coordinates of \object{Mrk\,817} ($l$=100.3\degree, $b$=53.5\degree),
using the dynamical definition of the LSR.
For the reference \hi\ column density in Complex C in the \object{Mrk\,817} direction, we use the value
log\,$N$(\hi)=19.50$\pm$0.01 derived by \citet{shull2011} from a Green Bank Telescope (GBT) 21\,cm spectrum 
taken at 9.1\arcmin\ angular resolution. This closely matches the value log\,$N$(\hi)=19.51$\pm$0.55 obtained by 
summing the two Complex C components fit toward \object{Mrk\,817} by \citet{french2021} 
in their study of \hi\ Lyman-series absorption.

\section{UV Analysis} \label{sec:analysis}

\subsection{Voigt Profile Fitting} \label{subsec: vpfit}

\begin{figure*}[!ht]
\includegraphics[width=\textwidth]{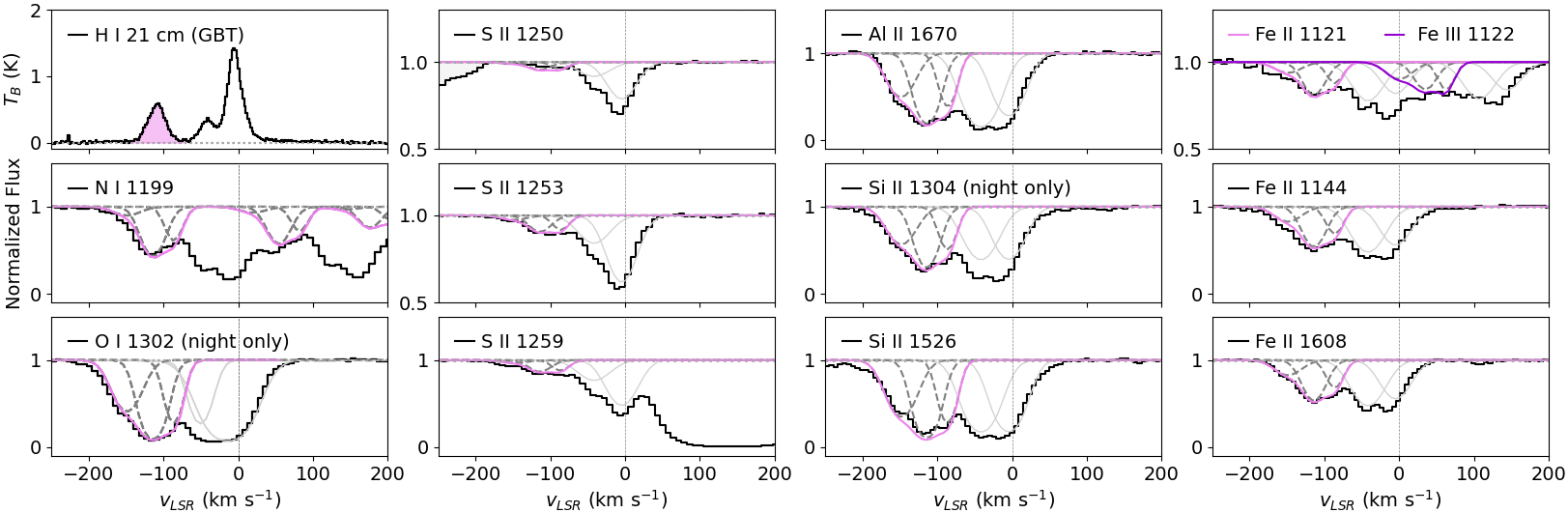}
\caption{Voigt profile fits to the low-ion absorption from Complex C in the coadded HST/COS spectrum of \object{Mrk\,817}. Normalized flux is plotted against LSR velocity for 11 low-ionization metal lines. A 21\,cm spectrum from the GBT is included in the top-left panel. The Complex C components are centered at $v_{\rm LSR}$=$-$149, $-$114, and $-86$\kms. The pink curves show the Complex C model adopted in this study. Individual components are shown with gray lines, dashed for Complex C and solid for Milky Way components.
The data have been binned by 3 pixels.}
\label{fig:fit}
\end{figure*}

We used the VPFIT software package version 12.2 \citep{carswell2014} to fit Voigt profiles 
to the MW and HVC absorption in 
\oi\ $\lambda$1302, \sw\ $\lambda\lambda$1253, \no\ $\lambda\lambda$1199.5,1200.2,1200.7, 
\few\ $\lambda$1121, 1144, 1608, \alw\ $\lambda$1670, and \siw\ $\lambda\lambda$1304. 
We focused on these six low-ionization species as they are well-suited to studies of the metallicity and 
dust depletion in Complex C. O, S, and N are relatively undepleted (volatile) elements \citep{jenkins2009}, 
whereas Fe, Si, and Al are refractory elements that deplete onto dust grains. 

For our fitting methodology, we normalized the COS spectra around each 
line of interest, using spline fits to regions of unabsorbed continuum. 
We assumed the component structure in the Milky Way and in Complex C was common to all low-ion lines under study,
so we tied the line centroids and line widths for the \oi, \no, \sw, \siw, \few, and \alw\ fits.
This assumption reflects the expected co-spatiality of these species given their similar ionization potentials, and allows
VPFIT to solve for the column density as the one free parameter for each ion in each component.
Our VPFIT models used the COS line spread function (LSF) for LP4 for the G130M/1222 and G160M/1533 settings. 
We fit the full profile of each ion, i.e. we include the MW (low-velocity) and Complex C (high-velocity) 
components in each fit.

We found evidence for three Complex C components and two MW components
in all low ions under study, 
with a strong Complex C component at $v_{\rm LSR}=-113.8\pm1.5$\kms\ 
surrounded by two weaker components at $-148.6\pm2.5$\kms\ and $-86.0\pm1.7$\kms. 
Our adopted VPFIT model to Complex C is therefore a three-component fit.
This model is shown in Figure~\ref{fig:fit} and the model parameters are given 
in Table~\ref{tab:columns}.
We calculate the total column density in Complex C in each ion by summing
across the three components. This is necessary because the reference \hi\ column density 
in Complex C is also summed across velocity components, so integrated metal measurements 
are needed for a like-to-like comparison.
We used VPFIT to calculate the weighted error on the total column density of each ion; 
this error is considerably less than the quadrature sum of the individual errors, 
because those errors are correlated due to degeneracies among the individual components.
This allows for precise calculation of the metallicities and depletions.

\begin{deluxetable*}{ll c c c c}[!ht]
\tabletypesize{\footnotesize}
\tablecaption{Complex C Absorption Parameters toward \object{Mrk\,817} (3-Component Models)$^a$}
\tablewidth{0pt}
\tablehead{\colhead{Ion} & \colhead{Lines Fit} & 
\colhead{log\,$N_1$} & 
\colhead{log\,$N_2$} & 
\colhead{log\,$N_3$} & 
\colhead{log\,$N_{\rm tot}$}\\
& & \colhead{($N$ in cm$^{-2}$)} & 
\colhead{($N$ in cm$^{-2}$)} & 
\colhead{($N$ in cm$^{-2}$)} & \colhead{($N$ in cm$^{-2}$)}}
\startdata
\oi\   & $\lambda$1302                 & 
14.371$\pm$0.062 & 
$>$14.715$^b$ & 
14.319$\pm$0.129 & 15.72$^{+0.24\,b}_{-0.16}$\\  
\sw\   & $\lambda$1253$^c$                 &  
13.512$\pm$0.162 & 
14.017$\pm$0.081 & 
13.834$\pm$0.119 & 14.311$\pm$0.033\\
\no\   & $\lambda\lambda$1199.5,1200.2,1200.7 & 
13.048$\pm$0.143 & 
13.867$\pm$0.057 & 
13.508$\pm$0.119 & 14.068$\pm$0.013\\ 
\siw\  & $\lambda$1304$^d$           & 
13.912$\pm$0.021 & 
14.136$\pm$0.017 & 
13.778$\pm$0.029 & 14.445$\pm$0.009\\
\few\  & $\lambda\lambda$1121,1144,1608  & 
13.516$\pm$0.074 & 
13.953$\pm$0.063 &
13.612$\pm$0.126 & 14.213$\pm$0.014\\ 
\fet\ & $\lambda$1122 &
13.377$\pm$0.175 &
13.518$\pm$0.134 &
13.388$\pm$0.150 & 13.910$\pm$0.143$^e$\\ 
\alw\  & $\lambda$1670                 &  
12.597$\pm$0.062 &  
12.891$\pm$0.068 & 
12.520$\pm$0.133 & 13.177$\pm$0.015\\ 
\enddata
\label{tab:columns}
\tn{a}{Our VPFIT models reveal three components within Complex C. 
We use the same component structure for all seven low ions, with 
$v_1=-148.6\pm2.5$\kms, $b_1=24.6\pm1.8$\kms,
$v_2=-113.8\pm1.5$\kms, $b_2=20.4\pm2.0$\kms, 
$v_3=-86.0\pm1.7$\kms, and $b_3=15.2\pm2.0$\kms.
We present the column densities of each component and the total column density summed 
across all three. The errors on the summed totals are the weighted uncertainties from 
VPFIT, and are much less than the quadrature sum of the individual errors.} 
\tn{b}{For \oi\ $\lambda$1302, the COS data only provide a lower limit for the total column density, because of saturation in component 2. Instead we adopt the value derived by \citet{collins2003} 
from unsaturated \oi\ absorption in \emph{Far Ultraviolet Spectroscopic Explorer} data.}
\tn{c}{\sw\ $\lambda$1250 is too weak and $\lambda$1259 is too blended to be included in the fit.}
\tn{d}{\siw\ $\lambda\lambda$1260, 1190, 1193, and 1526 are saturated in Complex C and are 
not included in the fit.}
\tn{e}{Our error budget on $N$(\fet) includes an additional systematic error of 0.13\,dex to reflect the uncertainty in deblending the Complex C absorption in \fet\ 1122 from low-velocity absorption in \few\ 1121.}
\end{deluxetable*}

\subsection{\textsc{Cloudy} Photoionization Models} \label{subsec:cloudy}

An essential step in metallicity and dust depletion analysis is the calculation of 
ionization corrections (ICs). ICs account for the amount of each ion hidden in unseen 
ionization stages. They are defined as the difference between the true elemental abundance 
(which we seek) and the observed ion abundance (which we measure). 
In a photoionization model, the true abundance is known, so the IC can be calculated as 

\begin{equation}
    \mathrm{IC}(\mathrm{X}^i)=\mathrm{[X/H]_{model}}-[\mathrm{X}^i/\textsc{H i}],
\end{equation}

where the square brackets denote logarithmic abundances relative to solar, i.e. 
[X/H]=(log\,$N_{\rm X}$--log\,$N_{\rm H}$)--log\,(X/H)$_\odot$ and
[X$^i$/\hi]=(log\,$N_{\rm X^{i}}$--log\,$N_{\rm H\,I}$)--log\,(X/H)$_\odot$.
We adopt solar (photospheric) abundances from \citet{asplund2009}, namely
log\,(O/H)$_\odot$=$-$3.31, log\,(S/H)$_\odot$=$-$4.88, log\,(N/H)$_\odot$=$-$4.17,
log\,(Si/H)$_\odot$=$-$4.49, log\,(Fe/H)$_\odot$=$-$4.50, and log\,(Al/H)$_\odot$=$-$5.55.

We ran customized photoionization models using the radiative transfer
code \textsc{Cloudy} version 22.01 \citep{ferland2017} to calculate 
the ICs in Complex C for each element under study (H, O, S, N, Si, Al, and Fe). 
Our modeling methodology is based on the procedures described in \citet{cashman2022}. 
In brief, we assume the cloud is a plane-parallel slab exposed to the 3D MW radiation field, 
with the SED taken from \citet{fox2005} and the normalization at the distance of Complex C from 
\citet{blandhawthorn2019}. We ran a grid of models at different gas 
density, and found the solution that best matches the observed ratio of 
log\,[$N$(\ion{Fe}{3})/$N$(\ion{Fe}{2})]=$-$0.30$\pm$0.13 in Complex C. 
This yielded log\,($n_{\rm H}$/cm$^{-3}$)=$-$1.47$\pm$0.13. 
Alternative methods to solve for the gas density 
are to use the \ion{C}{2}$^*$/\ion{C}{2} or \ion{Si}{3}/\ion{Si}{2} ratios.
We cannot derive a reliable density from \ion{C}{2}$^*$/\ion{C}{2} as 
\ion{C}{2} $\lambda$1334 is heavily saturated in Complex C, but
the density derived from the \ion{Si}{3}/\ion{Si}{2} 
ratio is log\,$(n_{\rm H}$/cm$^{-3})\approx-$1.45, 
very close to the density derived from \ion{Fe}{3}/\ion{Fe}{2}. 
Our derived density is close to the Complex C density 
of log\,$(n_{\rm H}$/cm$^{-3})=-$1.30 derived by \citet{shull2011}
assuming the cloud has a similar line-of-sight depth as angular extent.
We ran the \textsc{Cloudy} models using the total metal column densities summed 
across the three Complex C components,
since our \hi\ measurement is integrated in a similar way. In principle, we could run the models
on the individual components, but this would involve assumptions on how to split up the \hi,
so instead we present an integrated model.

The ICs are shown as a function of density for each of the ions under study in the top panel 
of Figure~\ref{fig:results}. Our adopted IC for each ion is calculated at the best-fit density.
The IC is very small (and flat) for \oi\ and \no\ (0.01--0.03 dex level), 
moderate for \few\ ($\approx$0.15 dex level), 
and at the $\approx$0.20--0.25\,dex level for \siw, \sw, and \alw.

\subsection{Chemical Abundances} \label{subsec:abundances}

\begin{deluxetable}{lcccc}[!ht]
\tablecaption{Complex C Abundances and Depletions}
\tabletypesize{\footnotesize}
\tablewidth{0pt}
\tablehead{\colhead{Ion} & \colhead{[X$^i$/\ion{H}{1}]$^a$} & \colhead{IC(X$^i$)$^b$} &  
\colhead{[X/H]$^c$} & \colhead{$\delta_\mathrm{S}$(X)$^d$ }}
\startdata
\multicolumn{3}{l}{\bf Volatile elements}\\
\oi\   & $-$0.47$^{+0.28}_{-0.22}$ & $-$0.01$\pm$0.01 & $-$0.48$^{+0.28}_{-0.22}$ & +0.03$^{+0.24}_{-0.17}$ \\ 
\sw\   & $-$0.31$\pm$0.15 & $-$0.20$\pm$0.03 & $-$0.51$\pm$0.16 & ... \\
\no\   & $-$1.26$\pm$0.15 & +0.03$\pm$0.01 & $-$1.23$\pm$0.15 & $-$0.72$\pm$0.05 \\
\multicolumn{3}{l}{\bf Refractory elements}\\
\siw\  & $-$0.57$\pm$0.15 & $-$0.24$\pm$0.03 & $-$0.80$\pm$0.15 &  --0.29$\pm$0.05 \\
\few\  & $-$0.79$\pm$0.15 & $-$0.14$\pm$0.07 & $-$0.93$\pm$0.17 &  --0.42$\pm$0.08 \\
\alw\  & $-$0.77$\pm$0.15 & $-$0.27$\pm$0.06 & $-$1.04$\pm$0.16 &  --0.53$\pm$0.08 \\
\enddata
\label{tab:abund}
\tn{a}{Ion abundance [X$^i$/\hi]=log [$N$(X$^i$)/$N$(\hi)]--log (X/H)$_\odot$ where X$^i$ is the observed ion of element X. The error is dominated by
beam-smearing uncertainties. We take the value of log\,$N$(\hi) in 
Complex C in the \object{Mrk\,817} direction as 19.50$\pm$0.01 \citep{shull2011}.}
\tn{b}{Ionization correction IC(X$^i$)=[X/H]$_\mathrm{model}$--[X$^i$/\hi], derived from \textsc{Cloudy} modeling (see Figure~\ref{fig:results}).}
\tn{c}{Gas-phase elemental abundance [X/H]=[X$^i$/\ion{H}{1}]+IC(X$^i$).} 
\tn{d}{Depletion of element X relative to sulfur $\delta_\mathrm{S}$(X)$\equiv$[X/S]=[X/H]$-$[S/H]. The errors are smaller than the abundance errors because the depletions do not depend on $N$(\hi) and so do not include a beam-smearing error, but they do include the uncertainties on X$^i$, \sw, IC(X$^i$), and IC(\sw).}
\end{deluxetable}

\begin{figure}[!ht] 
\includegraphics[width=0.47\textwidth]{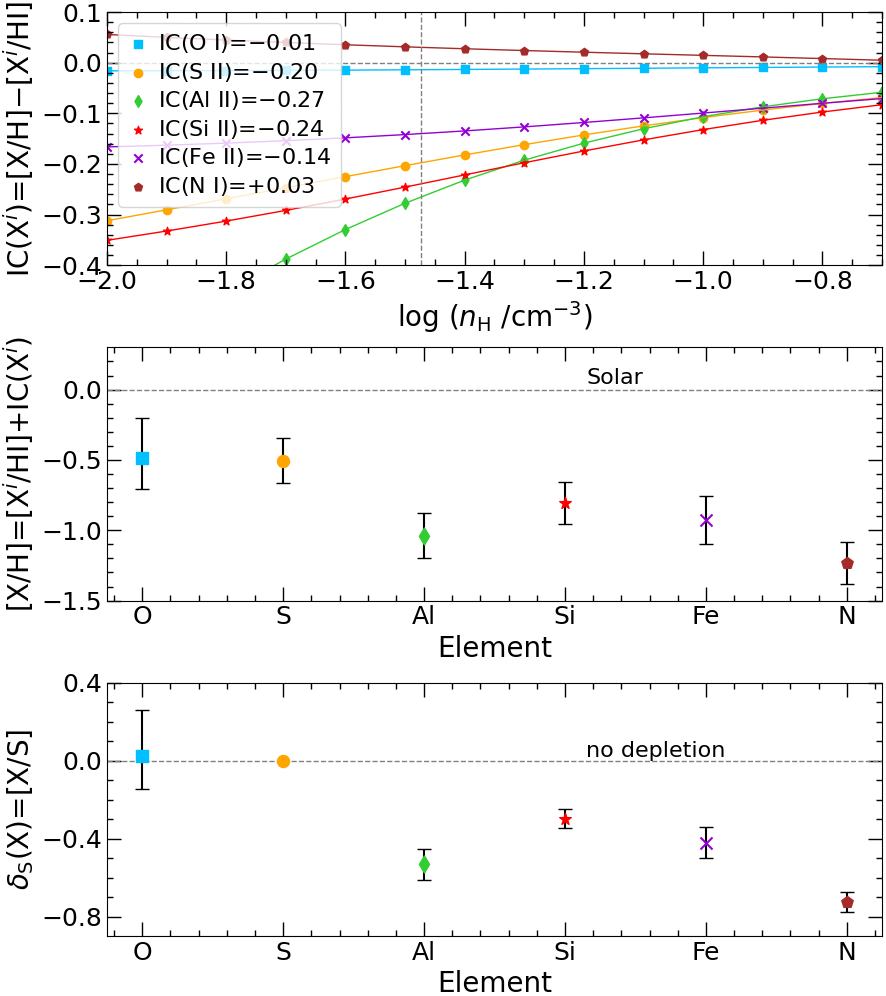}
\caption{Derivation of elemental abundances and dust depletions in Complex C toward \object{Mrk\,817}. 
The {\bf top panel} shows the ionization corrections (ICs) from our \textsc{Cloudy} models for 
six ions as a function of gas density. The best-fit density (dotted line) is derived by 
matching the observed \ion{Fe}{3}/\ion{Fe}{2} ratio in Complex C. The {\bf middle panel} shows 
the ionization-corrected abundances  of each element relative to hydrogen [X/H]. The 
{\bf bottom panel} shows the depletion of each element relative to sulfur [X/S], where S is 
chosen as an undepleted reference element. All data points are from our new COS analysis
except oxygen, which is based on the \ion{O}{1} measurement 
from \cite{collins2003} (see Table~\ref{tab:columns}).}
\label{fig:results}
\end{figure}

The gas-phase abundance of element X relative to hydrogen is

\begin{equation}
    [\text{X/H}]=[\text{X}^i/\textsc{H i}]+\mathrm{IC(X^i)}.
\end{equation}

The Complex C abundances calculated with Equation (2) are shown in the middle panel 
of Figure~\ref{fig:results} and listed in Table~\ref{tab:abund}.
For the volatile (undepleted) elements we find total abundances
across the three components of
[S/H]=$-$0.51$\pm$0.16, [O/H]=$-$0.48$^{+0.28}_{-0.22}$, 
and [N/H]=$-$1.23$\pm$0.15,
where the errors include a 0.15\,dex systematic error to account for the beam-size 
mismatch between the UV measurements and the \hi\ emission measurements \citep{wakker2011},
and where the oxygen abundance uses the \citet{collins2003} value for $N$(\oi), 
since it is derived from unsaturated lines in the FUSE spectrum
(whereas the \oi\ $\lambda$1302 line is saturated).
These abundances agree with previous measurements along the \object{Mrk\,817} sightline 
\citep{collins2003, collins2007, shull2011} and other Complex C sightlines
\citep{richter2001, gibson2001, tripp2003}, and confirm that Complex C is a low-metallicity cloud
with a significant nitrogen under-abundance. 

\subsection{Dust Depletion Levels} \label{subsec:depletions}

Following convention, we define the depletion $\delta_\mathrm{S}$(X) of each refractory 
element X (Si, Al, or Fe) as the abundance of that element relative to sulfur:

\begin{equation}
    \delta_\mathrm{S}(\text{X}) \equiv [\text{X/S}] = [\text{X/H}] - [\text{S/H}].
\end{equation}

In principle, the depletions could be calculated relative to O, which has as a smaller 
ionization correction than S, and our results would be essentially unchanged if we did this, since
our values of [S/H] and [O/H] are identical within their errors (see Table~\ref{tab:abund}).
However, we adopt S as the reference element as 
the \sw\ column densities (derived from the high S/N COS spectrum) have much smaller error bars 
than the \oi\ column densities. 
The depletions for Complex C are plotted in the bottom panel of Figure~\ref{fig:results}
and presented in Table~\ref{tab:abund}.
Fe, Si, and Al all show similar levels of depletion, with
[Fe/S]= $-$0.42$\pm$0.08, [Si/S]= $-$0.29$\pm$0.05, and [Al/S]=$-$0.53$\pm$0.08. 
We also derive [N/S]=$-$0.72$\pm$0.05, though 
this indicates an intrinsic nitrogen under-abundance, not a dust effect, since N is a volatile element.

The depletions are derived purely from measurements made along a single UV sight line, and so 
do not depend on the beam-smearing errors that affect HVC metallicities derived from combining 
UV absorption with \hi\ emission measurements made over a much larger beam. 
Our Complex C depletion measurements are therefore much more precise than the metallicity measurements.

\section{Discussion and Conclusions}\label{sec:discussion}

Our analysis of the UV absorption in the high-S/N coadded COS spectrum of \object{Mrk\,817} has yielded sub-solar abundance ratios in Complex C of 
[Fe/S]= $-$0.42$\pm$0.08, [Si/S]=$-$0.29$\pm$0.05, and [Al/S]=$-$0.53$\pm$0.08. 
Since S is relatively undepleted but Fe, Si, and Al are known to deplete in the ISM
\citep[][]{savage1996, jenkins2009, konstantopoulou2022}, 
these ratios quantify the amount of dust depletion for Fe, Si, and Al.
These are the first significant detections of dust depletion in Complex C,
though limits on dust depletion were derived by \citet{tripp2022}.
As a guide, 0.5\,dex of Al depletion means that 
only one Al atom out of three is in the gas phase, with the remaining two in dust.
Similarly, two in five Fe atoms are in gas, and one in two Si atoms.
The majority of the metal mass in Complex C in these three elements is therefore locked into dust grains.

It is worth considering whether other, non-dust-related effects could explain the 
anomalous abundance ratios derived in Complex C. In principle, non-solar abundance 
ratios can be caused by nucleosynthetic effects, 
in which different elements have different production pathways in stars.
Indeed, this is the favored explanation for the low N/S ratio \citep{richter2001}, which is interpreted as evidence for an intrinsic under-abundance of N caused by its 
different origin than the $\alpha$-elements, 
rather than a dust effect, since N does not deplete \citep{jenkins2009}. 
However, for $\alpha$-to-$\alpha$ ratios like Si/S and Si/O, the intrinsic ratio is 
expected to be solar, which is why the observed sub-solar Si/S implies dust depletion. 
For the non-$\alpha$ elements Al and Fe, the picture is more complex, 
as Al is a light odd-$Z$ element and 
Fe is an Fe-peak element, so we cannot rule out that their under-abundance relative 
to S has a contribution from nucleosynthetic effects as well as from dust grains,
but we follow the standard ISM interpretation of this as a depletion effect.

The depletions of Fe, Si, and Al constitute clear evidence for dust 
in Complex C. The presence of dust was considered in earlier UV analyses of the 
Mrk 817 sightline
\citep{collins2003, collins2007, shull2011} and the nearby sightlines to
Mrk\,876 \citep{murphy2000}, PG\,1259+593 \citep{richter2001}, and 3C\,351 \citep{tripp2003}, 
but could not be confirmed owing to the lower S/N of the 
spectra available at the time. A search for thermal IR emission from dust in Complex C
found a tentative detection \citep{miville2005}, but this is controversial  \citep{peek2009}.

Beyond Complex C, very few detections of dust in HVCs exist in general.
UV dust depletion has been reported in the Magellanic Stream \citep{fox2013} and its 
Leading Arm \citep{richter2018},
and the outer-arm HVC \citep{tripp2012, tripp2022}, 
but all other UV searches for dust in HVCs have found little or no depletion
\citep{richter2009, richter2017b, cashman2022}.
Furthermore, searches for far-IR emission from dust in HVCs 
have generally produced non-detections 
\citep{wakker1986, bates1988, lenz2016, williams2012}, with the exception
of a possible detection in Complex M \citep{peek2009}. 

The presence of dust in Complex C argues against a purely extragalactic origin, despite the 
cloud's low metallicity, because dust grains originate in stars, 
predominantly in AGB stars and supernovae \citep[][]{ferrarotti2006, dwek2011}.
While detailed origin models for Complex C are out of the scope of this Letter, 
a successful model must now
simultaneously explain the cloud's low metallicity \emph{and} moderate dust depletion.
This combination could be explained by a ``polluted accretion" model where an infalling extragalactic cloud is enriched by Galactic dust grains. 
However, it is also consistent with a fountain-driven circulation model
\citep{fraternali2015, fraternali2017}, in which a dust-bearing cloud 
originating in the Galactic disk was driven into the halo and 
precipitated the cooling of the (lower metallicity) 
hot corona via thermal instabilities 
\citep[see also][]{maller2004, marasco2022, heitsch2022}.
In either of these models, Complex C is neither
neither purely Galactic nor purely extragalactic, but a combination of the two. 
Another possible origin is tidally stripped gas from satellite galaxies, such as
the Magellanic Clouds, which could potentially explain the low metallicity and the presence of dust.
In any of these scenarios, the gaseous assembly history of 
the Milky Way is complex, with gas flows exchanging metals and dust grains 
in the circumgalactic medium. 

In closing, our results in this Letter illustrate how very-high-S/N UV 
spectroscopy of AGN enables high-precision measurements of MW halo clouds. 
Such deep UV spectroscopy offers a range of scientific benefits.\\

{\it Acknowledgements.} 
We thank the referee for a constructive report.
Support for program 16196 was provided by NASA through a grant from the Space Telescope Science Institute, 
which is operated by the Association of Universities for Research in Astronomy, Inc., under NASA contract NAS5-26555. 
The COS data are available at \url{https:doi.org//10.17909/n734-k698}.\\
\\
{\it Facilities:} HST/COS \citep{green2012}.\\   
\\
{\it Software:} VPFIT \citep{carswell2014}, \textsc{Cloudy} \citep{ferland2017}.

\bibliographystyle{aasjournal}
\bibliography{ref}{}

\end{document}